\documentclass[conference]{IEEEtran}

\usepackage[noadjust]{cite}

\usepackage{booktabs}
\usepackage{multirow}
\usepackage{algorithm}
\usepackage[noend]{algpseudocode}

\usepackage{graphicx}
\usepackage[caption=false]{subfig}

\usepackage{tabularx}
\usepackage{amssymb}
\usepackage[usenames, dvipsnames]{color}
\usepackage{soul}

\usepackage{lipsum}
\usepackage{diagbox}

\usepackage{pifont}

\newcommand{\xmark}{\ding{55}}

\usepackage[normalem]{ulem}
\usepackage{geometry}
\renewcommand{\baselinestretch}{0.95}

\usepackage{tikz}
\newcommand\copyrighttext{\scriptsize 
\textcopyright~2020 IEEE. Personal use of this material is permitted. Permission from IEEE must be obtained for all other uses, in any current or future media, including reprinting/republishing this material for advertising or promotional purposes, creating new collective works, for resale or redistribution to servers or lists, or reuse of any copyrighted component of this work in other works. \\
Accepted to be Published in IEEE VLSI Test Symposium (VTS) 2020}
\newcommand\copyrightnotice{%
\begin{tikzpicture}[remember picture,overlay]
\node[anchor=south,yshift=0pt] at (current page.south) {\fbox{\parbox{\dimexpr\textwidth-\fboxsep-\fboxrule\relax}{\copyrighttext}}};
\end{tikzpicture}%
}

\begin{document}
\bstctlcite{IEEEexample:BSTcontrol}

\title{\LARGE DFSSD: Deep Faults and Shallow State Duality, A Provably Strong Obfuscation Solution for Circuits with Restricted Access to Scan Chain}

\author{
    \IEEEauthorblockN{
        Shervin~Roshanisefat\IEEEauthorrefmark{1}, 
        Hadi~Mardani~Kamali\IEEEauthorrefmark{1}, 
        Kimia~Zamiri~Azar\IEEEauthorrefmark{1}, 
        Sai~Manoj~Pudukotai~Dinakarrao\IEEEauthorrefmark{1}, \\
        Naghmeh~Karimi\IEEEauthorrefmark{3}, 
        Houman~Homayoun\IEEEauthorrefmark{2}, 
        Avesta~Sasan\IEEEauthorrefmark{1}}
    \IEEEauthorblockA{
        \IEEEauthorrefmark{1}Department of ECE, George Mason University, e-mail: \{sroshani, hmardani, kzamiria, spudukot, asasan\}@gmu.edu \\
        \IEEEauthorrefmark{3}Department of CSEE, University of Maryland, Baltimore County, e-mail: nkarimi@umbc.edu \\            
        \IEEEauthorrefmark{2}Department of ECE, University of California, Davis, e-mail: hhomayoun@ucdavis.edu \\
    }
}

\maketitle

\begin{abstract}
In this paper, we introduce DFSSD, a novel logic locking solution for sequential and FSM circuits with a restricted (locked) access to the scan chain. DFSSD combines two techniques for obfuscation: (1) \underline{\textbf{D}}eep \underline{\textbf{F}}aults, and (2) \underline{\textbf{S}}hallow \underline{\textbf{S}}tate \underline{\textbf{D}}uality. Both techniques are specifically designed to resist against sequential SAT attacks based on bounded model checking. The shallow state duality prevents a sequential SAT attack from taking a shortcut for early termination without running an exhaustive unbounded model checker to assess if the attack could be terminated. The deep fault, on the other hand, provides a designer with a technique for building deep, yet key recoverable faults that could not be discovered by sequential SAT (and bounded model checker based) attacks in a reasonable time.
\end{abstract}

\copyrightnotice


\section{Introduction}\vspace{-1mm}
To reduce the cost of semiconductor fabrication and shorten the time to market of integrated circuits (IC), most of the fabrication processes are pushed offshore \cite{yeh2012trends}. This globalization of supply chain has tremendously raised security concerns such as the possibility of third-party intellectual property (3PIP) theft, IC overproduction, Trojan insertion, and adversarial reverse engineering. To overcome such threats, various active and passive \emph{design-for-trust} mechanisms have been proposed in the literature, among which logic locking, \emph{a.k.a.} hardware obfuscation, has been manifested as proactive protection against all these threats \cite{azarthreats, 242034}. The validity and strength of the state-of-the-art logic locking solutions to protect IPs/ICs against adversaries in the manufacturing supply chain was seriously challenged in recent years after the introduction of the Boolean satisfiability attack (SAT Attack) \cite{subramanyan2015evaluating,el2015integrated,8474189}. After introduction of the SAT attacks, researchers investigated a body of locking solutions with the objective of resisting the SAT attack \cite{yasin2017provably,xie2017delay,shervintvlsi2020,8429401,kamali2019full,kolhecustom}. However, further research revealed that increasing resistance against SAT attack makes such solutions vulnerable against alternative (and even simpler) attack solutions such as Signal Probability Skew (SPS) and structural analysis-based attacks \cite{7858346,8715163,azar2019smt}. 

\begin{figure}
    \centering
    \includegraphics[width=0.69\columnwidth]{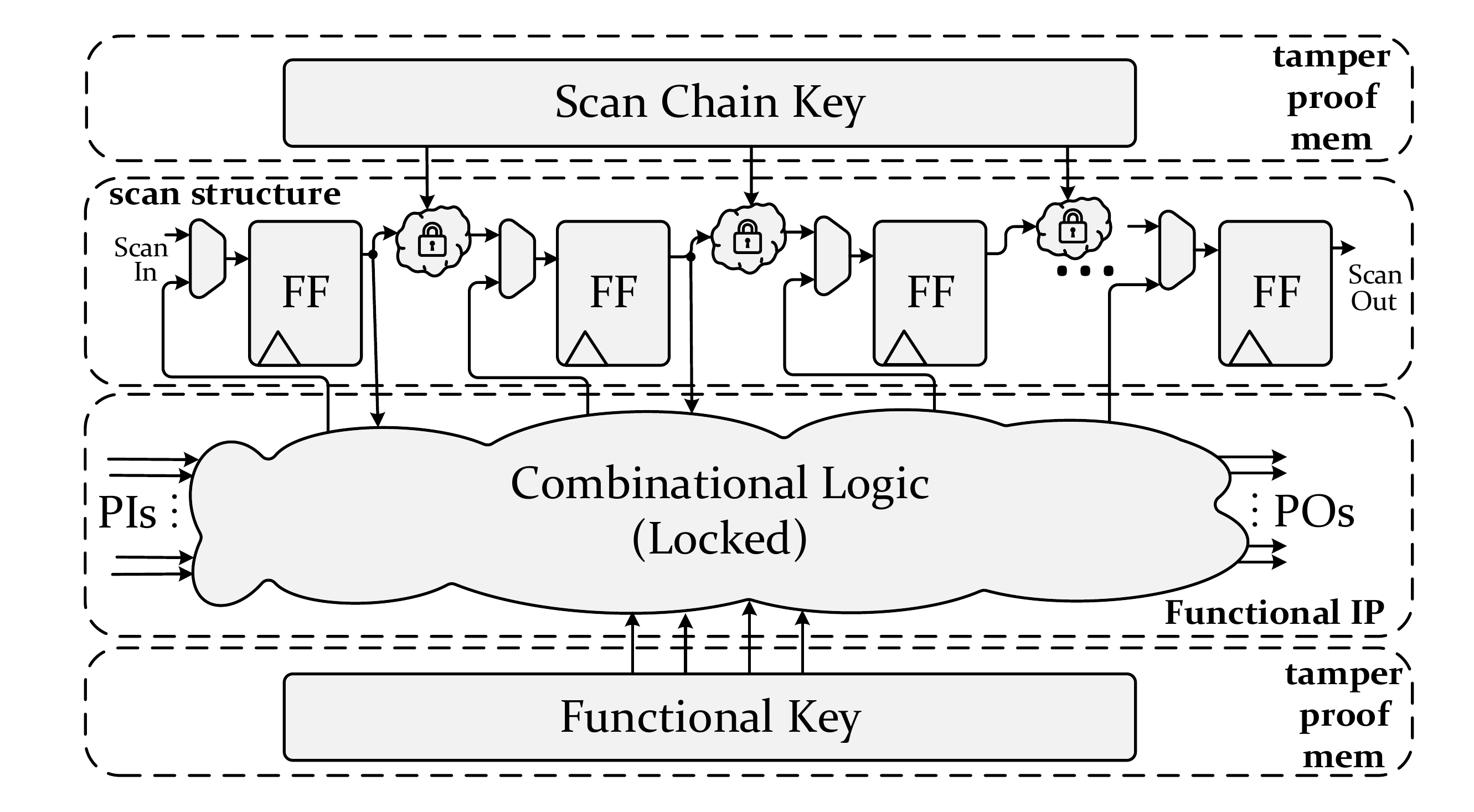}
    \caption{An obfuscated IC with restricted access to scan chain.}
    \label{dist_keys}
\end{figure}

The original SAT attack was only applicable to combinational circuits. However, the existence of the scan chain, allows an adversary to treat the FSM and sequential circuits as a combinational circuit; using the scan chain, the attacker to load desired input into scan registers, carry the attack for one cycle, and readout the output through the scan chain \cite{subramanyan2015evaluating}. Hence, to prevent the SAT attack on obfuscated sequential and FSM solutions, various means for restricting access to the scan chain \cite{8290974,DBLP:journals/corr/abs-1906-07806} was investigated. In this approach, which is illustrated in Fig. \ref{dist_keys}, an obfuscation solution is constructed using two key values: (1) a key for obfuscating the functional logic, and (2) a key for obfuscating the scan chain.

Restricting access to (or locking of) the scan chain, however, did not stop the researchers from developing variants of SAT attack solution capable of attacking an obfuscated circuit. Lack of access to the scan chain was addressed in~\cite{el2017reverse} by changing the attack model to find a sequence of inputs (rather than a single input) resulting in incorrect output. This attack, so-called unrolling-based SAT (UB-SAT) attack, expands the given FSM in time to be able to find a sequence of distinguishing inputs. 

To defend against UB-SAT and model checker based attacks in design with restricted access to the scan chain, in this paper, we introduce a new obfuscation solution denoted as Deep Faults and Shallow State Duality (DFSSD). The DSFFD obfuscation scheme exploits the weaknesses of the existing attacks in obfuscating FSM and sequential circuits and prevents these attacks from satisfying their early exit conditions, forcing them to become unbounded.  To build the DFSSD solution, we propose a combination of two concepts: (1) encrypting \textbf{Deep Faults (DF)}, the discovery of which requires specific traversal patterns with a large enough depth that cannot be reached by bounded model checkers or unrolling based SAT attacks. (2) encoding \textbf{Shallow State Duality (SSD)}, in which by implementing key-controlled duplicate states, the early termination conditions of the UB-SAT are violated. 

\section{Preliminary Background} \label{background} \vspace{-1mm}
As described earlier, limiting access to the scan chain removes the ability of the attacker to deploy a pure SAT attack on the combinational logic between internal scan registers, and has to revert to the weaker variant of SAT attacks such as UB-SAT (working with only primary input and primary output). Following is a short background on Scan chain obfuscation and proposed attack solutions for de-obfuscating such solutions needed for understanding the DFSSD.

\textbf{Securing Scan Chain Structure:} Several methods have been recently proposed in the literature to obfuscate the scan chains~\cite{karmakar2018encrypt,8105900}. To secure the test and debug operations, \cite{8290974}~proposed a design-for-security (DFS) flow that deploys a structure, denoted as \textit{Secure Cell} (SC). However, SC was compromised via the \textit{shift-and-leak} attack \cite{DBLP:journals/corr/abs-1906-07806}. Another early attempt in this domain was the \textit{Encrypt Flip-Flop} (EFF)~\cite{karmakar2018encrypt} scheme. In EFF the output of each scan flop is obfuscated based on a key value such that either the $Q$ or $Q_{bar}$ output is propagated in the scan chain, and accordingly, the \emph{scan-in} sequence is also modified. The EFF was also tackled by the \textit{ScanSAT} attack~\cite{alrahis2019scansat}. The \textit{Dynamically Obfuscated Scan} (DOS) \cite{8105900} scheme obfuscates the scan chain while periodically changing the obfuscation key during the test process. Assuming a hard to break scan chain obfuscation, the pure SAT attack could be no longer applied. Hence, an attacker should resort to SAT attack variants designed for attacking scan-access restricted obfuscation solutions by only relying on controllability (observability) of primary inputs (outputs). 

\textbf{Deobfuscation Methods Without Scan Chain Access:} El Massad et al.~\cite{el2017reverse} extended the SAT attack to circuits with no scan chain access, proposing an attack that only required access to the primary input/outputs of an activated chip. The attack procedure is shown in Algorithm \ref{SAT_algoritm}. Similar to the SAT attack, it has an iterative process for pruning the search space. However, due to the restricted access to the internal registers, rather than finding a Discriminating Input (in each iteration), it finds a sequence of inputs $X$ denoted as \emph{Discriminating Input Sequence} ($X_{DIS}$) that can generate two different outputs for the same input sequence for two different keys. In this algorithm, $C(X,K,Y)$ refers to the obfuscated circuit producing output sequence $Y$ using input sequence $X$ and key vector $K$, and $C_{BlackBox}(X)$ refers to the output sequence of the activated circuit for the same input sequence. After transforming the obfuscated circuit to a circuit SAT ($Model$) problem, the attack instantiates a Bounded Model Checker (BMC) to find the $X_{DIS}$. After the discovery of each $X_{DIS}$, the $Model$ is updated with a new condition to make sure that the next onset of keys, that will be discovered in the subsequent attack iterations, produce the same output for previously discovered $X_{DIS}$. This process continues until no further $X_{DIS}$ is found within the boundary of $b$. 

\begin{algorithm}[t]
\caption{\small Sequential Attack on Obfuscated Circuits \label{SAT_algoritm}}
\begin{algorithmic}[1]
\scriptsize
\State $b = initial\_boundary$, $Terminated = False$;
\State $Model = C(X,K_1,Y_1) \wedge C(X,K_2,Y_2) \wedge (Y_1 \ne Y_2)$;
\While {not $Terminated$}
    \While {$(X_{DIS},K_1,K_2)\leftarrow BMC(Model, b)=T$} 
        \State $Y_f \leftarrow C_{BlackBox}(X_{DI})$;
        \State $Model =\wedge ~C(X_{DIS},K_1,Y_f) \wedge C(X_{DIS},K_2,Y_f)$;
    \EndWhile

    \If{$\textbf{UC}(Model,b) \lor \textbf{CE}(Model, b) \lor \textbf{UMC}(Model)$} 
        \State $Terminated$;
    \EndIf
    \State $b=b+boundary\_step$;
\EndWhile
\end{algorithmic}
\end{algorithm}

After reaching the boundary, the algorithm checks three criteria to determine if the attack can be terminated: (1) \textbf{Unique Completion (UC):} This criterion checks for the uniqueness of the key. If there is only a single key that satisfying all previous \textit{DIS}es, the attack is terminated. (2) \textbf{Combinational Equivalence (CE):} If there is more than one key that agrees with all previously found $X_{DIS}$, the attack checks the combinational equivalency of the remaining keys. In this step, the input/output of FFs are considered as pseudo primary outputs/inputs allowing the attacker to treat the circuit as combinational. The resulting circuit is subjected to a SAT attack, and if the SAT solver fails to find a different output or next state for two different keys, it concludes that all remaining keys are correct and the attack terminates. (3) \textbf{Unbounded Model Check (UMC):} If UC and CE fail, the attack checks the existence of a DIS for the remaining keys using an unbounded model checker. This is an exhaustive search with no limitation on bound (or the number of unrolls). If no DIS is discovered, the existing set of DIS is a complete set, and the attack terminates. Otherwise, the bound is increased and previous steps are repeated. The original implementation of this attack \cite{el2017reverse} uses NuSMV as the model checker and is not scalable for larger circuits. Shamsi et al. improved this attack via implementing several tweaks in the attack procedure~\cite{8715053}. 

\section{Proposed Methods} \label{dfssd}
The practicality of UB-SAT attack (proposed in~\cite{el2017reverse}) is grounded on the use of a fast bounded model checker (BMC)~\cite{Clarke2012} and the implementation of early termination strategies to avoid the exhaustive search. This allows the attacker to avoid using time-consuming and exhaustive unbounded model checking runs for the discovery of DISes and to find the obfuscation key in a reasonable time. In this section, we describe an obfuscation solution that 1) prevent the UC and CE early termination, and 2) pushes the required bound for a BMC solver to an unreasonably large bound (which is defined at design time), resulting in unreasonable attack time against the proposed obfuscation solution.

\subsection{Shallow State Duality} \label{duplicated_states}
The first termination criterion (UC) relies on the uniqueness of the key and it fails if there is more than one valid key for the obfuscated circuit. In the sequential attack proposed in~\cite{el2017reverse}, UC was the main termination criterion for most of the benchmarks. For the second termination criteria (CE), successful termination relies on the equality of all next state and output values for remaining candidate keys for all input and state combinations. 

\begin{figure}
\begin{minipage}{\linewidth}
\centering
\includegraphics[width=0.78\columnwidth]{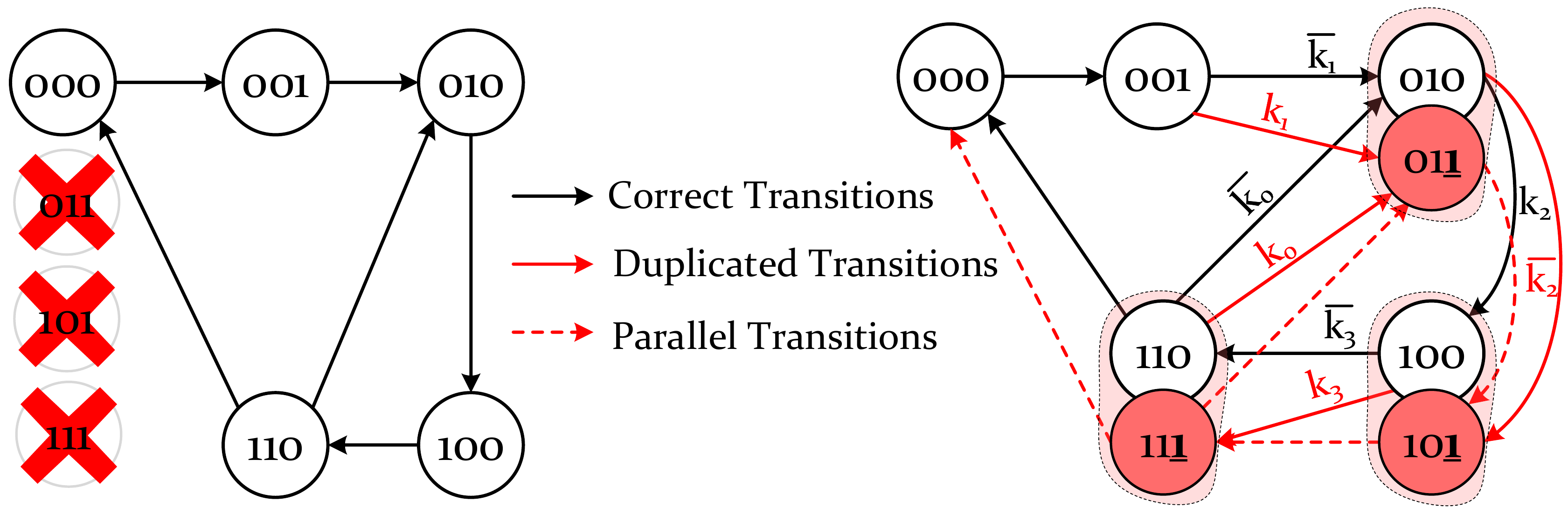}
\end{minipage}
\begin{minipage}{\linewidth}
\scriptsize
\centering
\setlength\tabcolsep{1.9pt}
\begin{tabular}{@{} c *{10}c @{}}
\cmidrule(lr){1-4} \cmidrule(lr){6-11}
Current & Next & Next & \multirow{3}{*}{output} & & Current & Next & Next & Next & Next & \multirow{3}{*}{output}  \\
State & State & State & & & State & State & State & State & State & \\
& I=0 & I=1 & & & & I,k$_i$=00 & I,k$_i$=10 & I,k$_i$=01 & I,k$_i$=11 \\
\cmidrule(lr){1-1} \cmidrule(lr){2-3} \cmidrule(lr){4-4} \cmidrule(lr){6-6} \cmidrule(lr){7-8} \cmidrule(lr){9-10} \cmidrule(lr){11-11}
000 & 001 & 001 & A & \textcolor{black}{$\longrightarrow$} & 000 & 001 & 001 & 001 & 001 & A \\
001 & 010 & 010 & B & \textcolor{black}{$\longrightarrow$} & 001 & 010 & 010 & \textcolor{red}{011} & \textcolor{red}{011} & B \\
010 & 100 & 100 & C & \textcolor{black}{$\longrightarrow$} & 010 & \textcolor{red}{101} & \textcolor{red}{101} & 100 & 100 & C \\
011 & \multicolumn{3}{l}{\xmark~NOT USED~\xmark} & \xmark & \textcolor{red}{011} & \textcolor{red}{101} & \textcolor{red}{101} & \textcolor{red}{101} & \textcolor{red}{101} & \textcolor{red}{C} \\
100 & 110 & 110 & D & \textcolor{black}{$\longrightarrow$} & 100 & 110 & 110 & \textcolor{red}{111} & \textcolor{red}{111} & D \\
101 & \multicolumn{3}{l}{\xmark~NOT USED~\xmark} & \xmark & \textcolor{red}{101} & \textcolor{red}{111} & \textcolor{red}{111} & \textcolor{red}{111} & \textcolor{red}{111} & \textcolor{red}{D} \\
110 & 000 & 010 & E & \textcolor{black}{$\longrightarrow$} & 110 & 000 & 010 & 000 & \textcolor{red}{011} & E \\
111 & \multicolumn{3}{l}{\xmark~NOT USED~\xmark} & \xmark & \textcolor{red}{111} & \textcolor{red}{000} & \textcolor{red}{000} & \textcolor{red}{011} & \textcolor{red}{011} & \textcolor{red}{E} \\

\cmidrule(lr){1-4} \cmidrule(lr){6-11}
\end{tabular}
\end{minipage}\hfill
\caption{(a) Original state transition graph, (b) Modified state transition graph with duplicate states (shallow state duality).}
\label{designtime}
\end{figure}

Our proposed solution for breaking both UC and CE termination checks is simply adding duplicate key controlled, yet valid states such that more than one valid key exists. We refer to this scheme as Shallow State Duality (SSD). This concept is illustrated in Fig. \ref{designtime}. In this example, the original FSM has five Reachable States (RS) and three Un-Reachable States (URS). In the modified FSM, the unreachable states are used to replicate three of the existing states such that the transition to the original or replicated state is controlled by a key. In this example, all the replicated states produce the same outputs as the original state and key bits are correct for both values of 0 and 1, although, it might not be the case in a different implementation. Therefore, the UC check fails as more than one correct key exists. In addition, in the CE check, the input to the registers is considered as a primary output. Hence, for duplicated states, two different key values do not generate the same output as they do not reach the same state. Note that the SSD is not a form of obfuscation as multiple keys are correct keys and it should be combined with our obfuscation solution which is described next. However, it is an effective and low-overhead technique to prevent early termination of the UB-SAT attack and its variants.

The duplicate states can be added during the state encoding (design time) or after logic synthesis (physical design time). Fig.~\ref{designtime} shows an example of a state transition graph encoded with duplicated~states at design time. The circuit functionality remains unchanged while the circuit has $2^4=16$~correct keys.  

\begin{algorithm}[t]
\caption{\small Extracting an unreachable state with minimum hamming distance from a reachable state \label{qbf_algoritm}}
\begin{algorithmic}[1]
\scriptsize
\State $boundary = limit$, $i = 1$, $hd = 1$
\State $Model = C_{comb}(X,S,O,S_{next}) \wedge C_{comb}(X_1,S_{init},O_1,S_1)$
\While {$true$}
    \State $A = (\forall (S, X), \exists S_{urs}, S_{next} \neq S_{urs}) \wedge (HD(S_{urs}, S_i) == hd)$
    \If{$(S_{i-1}, S_i, S_{urs})\leftarrow QBF(Model \wedge A)=T$}
        \State return $S_{i-1}, S_i, S_{urs}$
    \ElsIf{$i < boundary$}
        \State $i = i + 1$
        \State $Model =\wedge ~C_{comb}(X_i,S_{i-1},O_i,S_i)$
    \ElsIf{$i == boundary$ and $hd < output\_width$}
        \State $i = 1$, $hd= hd+1$
    \EndIf
\EndWhile
\end{algorithmic}
\end{algorithm}

For adding duplicate states to a synthesize netlist, we first need to find a few unreachable states ($S_{urs}$). The Algorithm \ref{qbf_algoritm} describes our approach for finding such states using a quantified Boolean formula (QBF) solver. To minimize the logic (overhead) needed for encoding the duplicate states, we search for $S_{urs}$ with minimum hamming distance (HD) from one of reachable (existing) states. In this Algorithm, inputs, states, outputs, and next states are defined as $X$, $S$, $O$, and $S_{next}$, respectively, and $C_{comb}$ refers to the combinational representation of the original circuit in which the input/output of FFs are considered as pseudo primary outputs/inputs (similar to CE check in UB-SAT attack). 

After initializing the boundary limit and defining the desired hamming distance (e.g. hd=1), a model consisting of two $C_{comb}$ instances is created. To find a $S_{urs}$, one instance of $C_{comb}$ is used as $C_{comb}(X,S,O,S_{next})$ with \textit{for-all} condition on its primary inputs ($X$) and current states ($S$) to generate all the outputs ($O$) and next states ($S_{next}$) that could be produced by the $C_{comb}$. By assuming $S_{next} \neq S_{urs}$, the QBF solver will attempt to find a set of values for $S_{urs}$ that is not a part of the generated $S_{next}$. Then, to select a URS from the set of $S_{urs}$ that has a hamming distance of $hd$ from a RS, another instance of $C_{comb}$ as $C_{comb}(X_1,S_{init},O_1,S_1)$ is used. In the QBF solver, this instance produces RSes ($S_1$) that are reachable from the initial state ($S_{init}$). Any URS in $S_{urs}$ with HD of one from the $S_1$ could be considered as the answer. If such a URS was not found, a new copy of $C_{comb}$ as $C_{comb}(X_2,S_{1},O_2,S_2)$ is added to the model to produce RSes that are reachable from the initial state in two cycles. If necessary, this unrolling continues until the boundary limit to check $S_{urs}$ with all RSes reachable in $i$ cycles. When a URS is found, it is added to the netlist by adding the logic to make the transition between the URS and the original states based on the key value. The URS will produce the same output as the original RS it was duplicated from and will transition to the same next state(s) (or the duplicate of the next states).

\subsection{Deep Faults}
The sequential attack \cite{el2017reverse} relies on a bounded search space for finding a discriminating input sequence $X_{DIS}$, and it keeps increasing the boundary if the termination checks fail. The $X_{DIS}$ is a sequence of inputs, each forces a transition to a new state until a discriminating state is reached, where a discriminating state refers to a state whose output is different for the same input with two different keys (a DIP condition). This state traversal (based on $X_{DIS}$) will not include any other discriminating state transition or repeated state. \textit{Such a \underline{discriminating state} could only be found if the \underline{shortest state traversal path} from the initial state to that state is \underline{shallower than the boundary} condition (the number of transitions) specified when invoking the BMC solver}.

The traversal depth of a sequential/FSM circuit is defined as the maximum number of state traversals (starting from initial state) where no state is visited twice. However, the sequential/FSM circuits may have a limited traversal depth~\cite{Mneimneh}. This makes a BMC a plausible attack for finding all possible DIS in such circuits, as all states can be visited within a reasonably small bound. Our solution to protect against BMC formulated attack (e.g., \cite{el2017reverse}) is to increase the traversal depth of the FSM/sequential circuits and push the impact of wrong keys into deep states beyond reach of the BMC (with reasonable bound). This makes the discovery of such DIS unreasonably time consuming. We refer to such faults as deep faults. 

\begin{figure}
    \centering
    \includegraphics[width=0.75\columnwidth]{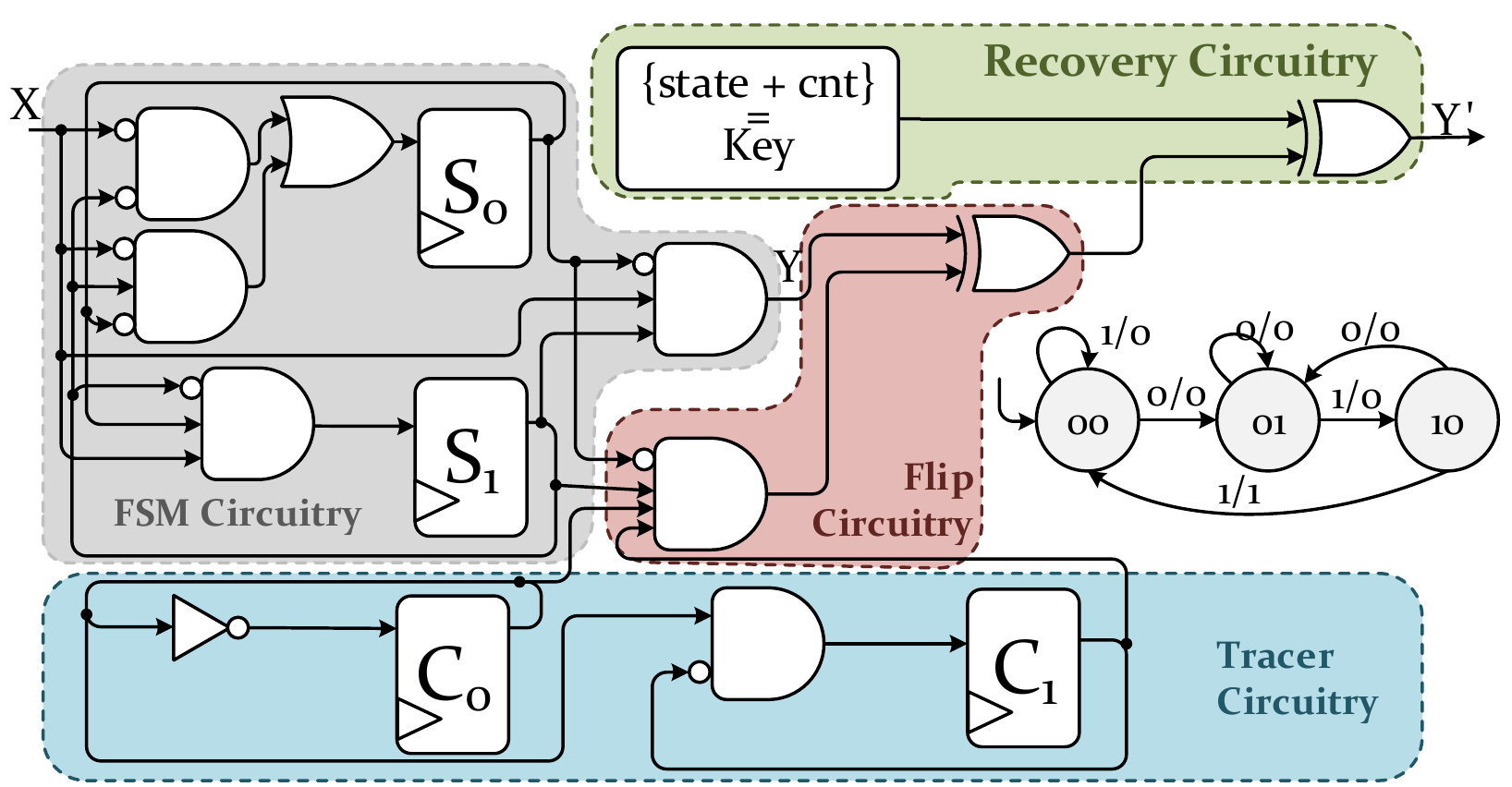}
    \caption{Implementation of Deep Fault (DF) obfuscation for a 011 detector. The protected pattern include the counter ($C_1C_0$) FFs, and the ($S_1S_0$) state FFs. The state transition graph of the circuit is shown in the middle right.}
    \label{deepfault}
\end{figure}

Our obfuscation methodology for creating Deep Faults (DF) is described via the example shown in Fig.~\ref{deepfault}. The circuit targeted for obfuscation is a simple `0-1-1' input sequence detector. As illustrated, the DF is implemented by adding 1) a \underline{tracer circuit}, 2) a \underline{flip circuit}, and 3) a \underline{recovery circuit} to the original circuit. The tracer, as described earlier, is a function-modified counter or a LSFR that changes its state each time a triggering event is observed. The triggering events can be selected state transitions, state visits, or simply the rising edge of the clock. For simplicity, in Fig. 4, the triggering event is the clock and the tracer is a 2-bit counter. The flip circuit toggles the value of a single primary output of the original circuit when a protected pattern is observed. The protected pattern is a predefined pattern generated by combining selected state registers (from the original design) and tracer's state register.  In Fig. \ref{deepfault}, the flip circuit is shown as a four-input AND gate that fires when the protected '10,11' pattern for '$S_1S_0,C_1C_0$' is observed. The last component of the Deep Fault obfuscation is the recovery circuit that toggles the output signal when the inserted obfuscation key agrees with the protected pattern. Hence, when the correct key is applied (1011 in this case), the previously flipped output related to the protected pattern will flip back (recovered) by the recovery circuit, however, insertion of a wrong key will result in flipping a correct output. 

Table \ref{truthtable} shows the truth table of the circuit in Fig. \ref{deepfault} for all key-combinations. In this circuit, when the DF is subjected to UB-SAT attack, each input can only rule out a single wrong key. Thereby, the pruning power of each discovered DIS is very limited, and the correct key is found only when the protected input is tested. This concept is similar to obfuscation solutions using point functions (e.g. SARLock \cite{yasin2016sarlock} and Anti-SAT \cite{xie2016antisat}). However, there is a fundamental difference. In point functions, the adversary uses a random input, and although the average case or worse case attack time is an exponential function of the key size, the attacker can potentially discover the correct key with a single lucky attempt. However, in deep faults, the discovery of DISes is conditioned on the tracer state, which cannot be directly controlled by input. Hence, it can guarantee a minimum bound on the number of required DISes before the discovery of the fault, which is at least equal to the number of cycles needed for the tracer to reach the fault generation state. This is a necessary condition for the generation of the fault, but it is not enough. For the fault to occur, the selection of state registers of the circuit that are included as a part of a protected pattern should also reach the fault generating pattern. Hence, the number of required DISes, which is equal to the number of cycles to reach the protected pattern, can be far~larger.  

\begin{table}[t]
\scriptsize
\centering
\caption{Truth table for the 011 detector obfuscated using a two-bit counter. $S_1S_2$ and $C_1C_2$ denote the state and counter FFs respectively. Y is the original circuit output and $K_n$ is the output for each four-bit key. The protected pattern and key are 1011. State values of 11xx are unreachable.}
\label{truthtable}
\setlength\tabcolsep{1.0pt} 
\setlength\extrarowheight{-1pt}
	\scalebox{0.9}{
\begin{tabular}{@{} c|c|cccccccccccccccc @{}}
\toprule 
$S_1S_2C_1C_2$ & Y & $K_0$ & $K_1$ & $K_2$ & $K_3$ & $K_4$ & $K_5$ & $K_6$ & $K_7$ & $K_8$ & $K_9$ & $K_{10}$ & $K_{11}$ & $K_{12}$ & $K_{13}$ & $K_{14}$ & $K_{15}$ \\ \midrule
0000 & \checkmark & \textcolor{red}{\xmark} & \checkmark & \checkmark & \checkmark & \checkmark & \checkmark & \checkmark & \checkmark & \checkmark & \checkmark & \checkmark & \checkmark & \checkmark & \checkmark & \checkmark & \checkmark \\
0001 & \checkmark & \checkmark & \textcolor{red}{\xmark} & \checkmark & \checkmark & \checkmark & \checkmark & \checkmark & \checkmark & \checkmark & \checkmark & \checkmark & \checkmark & \checkmark & \checkmark & \checkmark & \checkmark \\
0010 & \checkmark & \checkmark & \checkmark & \textcolor{red}{\xmark} & \checkmark & \checkmark & \checkmark & \checkmark & \checkmark & \checkmark & \checkmark & \checkmark & \checkmark & \checkmark & \checkmark & \checkmark & \checkmark \\
0011 & \checkmark & \checkmark & \checkmark & \checkmark & \textcolor{red}{\xmark} & \checkmark & \checkmark & \checkmark & \checkmark & \checkmark & \checkmark & \checkmark & \checkmark & \checkmark & \checkmark & \checkmark & \checkmark \\
0100 & \checkmark & \checkmark & \checkmark & \checkmark & \checkmark & \textcolor{red}{\xmark} & \checkmark & \checkmark & \checkmark & \checkmark & \checkmark & \checkmark & \checkmark & \checkmark & \checkmark & \checkmark & \checkmark \\
0101 & \checkmark & \checkmark & \checkmark & \checkmark & \checkmark & \checkmark & \textcolor{red}{\xmark} & \checkmark & \checkmark & \checkmark & \checkmark & \checkmark & \checkmark & \checkmark & \checkmark & \checkmark & \checkmark \\
0110 & \checkmark & \checkmark & \checkmark & \checkmark & \checkmark & \checkmark & \checkmark & \textcolor{red}{\xmark} & \checkmark & \checkmark & \checkmark & \checkmark & \checkmark & \checkmark & \checkmark & \checkmark & \checkmark \\
0111 & \checkmark & \checkmark & \checkmark & \checkmark & \checkmark & \checkmark & \checkmark & \checkmark & \textcolor{red}{\xmark} & \checkmark & \checkmark & \checkmark & \checkmark & \checkmark & \checkmark & \checkmark & \checkmark \\
1000 & \checkmark & \checkmark & \checkmark & \checkmark & \checkmark & \checkmark & \checkmark & \checkmark & \checkmark & \textcolor{red}{\xmark} & \checkmark & \checkmark & \checkmark & \checkmark & \checkmark & \checkmark & \checkmark \\
1001 & \checkmark & \checkmark & \checkmark & \checkmark & \checkmark & \checkmark & \checkmark & \checkmark & \checkmark & \checkmark & \textcolor{red}{\xmark} & \checkmark & \checkmark & \checkmark & \checkmark & \checkmark & \checkmark \\
1010 & \checkmark & \checkmark & \checkmark & \checkmark & \checkmark & \checkmark & \checkmark & \checkmark & \checkmark & \checkmark & \checkmark & \textcolor{red}{\xmark} & \checkmark & \checkmark & \checkmark & \checkmark & \checkmark \\
1011 & \checkmark & \textcolor{red}{\xmark} & \textcolor{red}{\xmark} & \textcolor{red}{\xmark} & \textcolor{red}{\xmark} & \textcolor{red}{\xmark} & \textcolor{red}{\xmark} & \textcolor{red}{\xmark} & \textcolor{red}{\xmark} & \textcolor{red}{\xmark} & \textcolor{red}{\xmark} & \textcolor{red}{\xmark} & \checkmark & \textcolor{red}{\xmark} & \textcolor{red}{\xmark} & \textcolor{red}{\xmark} & \textcolor{red}{\xmark} \\
11xx & \multicolumn{17}{c}{S=11 is unreachable} \\

\bottomrule
\end{tabular}
}
\end{table}

The lower bound for finding the protected pattern could be defined based on the tracer event counting mechanism. For simplicity, let's assume a counter is used as the tracer circuit. 

\textbf{Lemma 1.} \emph{The bound requirement for a BMC solver to find the protected pattern of a deep fault which is implemented using a simple clocked counter is $C=2^w$ where $w$ is width of the counter.}

\begin{figure}
    \centering
    \includegraphics[width=0.95\columnwidth]{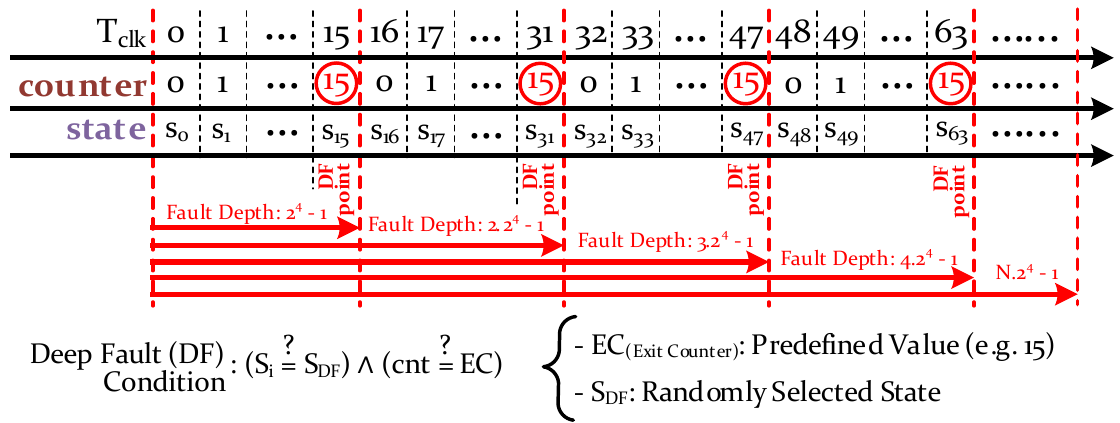}
    \caption{Deep faults mechanism for an always-on counter.}
    \label{dis_gen1}
\end{figure}

\emph{Proof.} The protected pattern consists of two parts: 1) w bits of tracer (counter) register bits, and s bits of state registers. As illustrated in Fig. \ref{dis_gen1}, the portion of protected pattern implemented by counter is reached every $C=2^w$ cycles. However, the state transition does not have a predefined traversal order, and the fault is only generated when the protected pattern (state-tracer) is observed. If the s bits of state registers, that are selected for inclusion in the protected pattern, do not take the needed value to build the protected pattern at cycle $C$, the fault is not generated. The next viable cycle for reaching the protected pattern will be at $2\times C$ or in general at $N \times C$. Hence, the minimum bound requirement for a BMC attack to discover the fault is $C=2^w$. $\blacksquare$

\textbf{Lemma 2.} \emph{The bound requirement for a BMC solver to find the protected pattern for a deep fault which is implemented using  a tracer that counts a selected \textbf{state transition} is $M+C\times L+Q$, where $M$ is the shortest path to reach the selected state transition from the initial state, $C=2^w$, $w$ is width of the counter, $L$ is the shortest sequence of state transitions to visit the selected triggering transition twice (shortest cycle including the triggering transition), and $Q$ is the number of state transitions to reach a state whose encoding completes the protected pattern signature.}

\begin{figure}
    \centering
    \includegraphics[width=0.95\columnwidth]{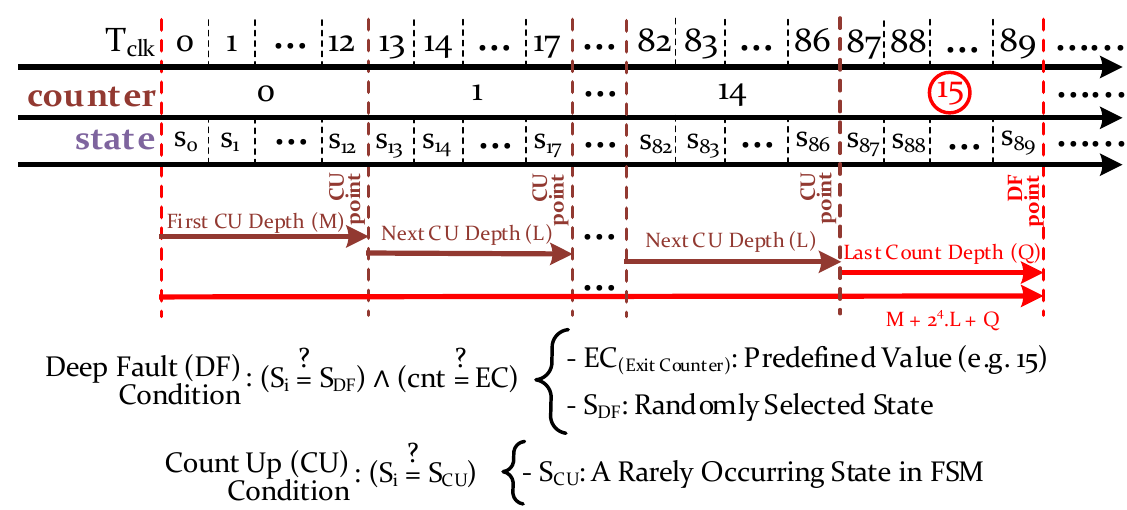}
    \caption{Deep faults mechanism for transition-triggered counter.}
    \label{dis_gen2}
\end{figure}

\emph{Proof.} As shown in Fig. \ref{dis_gen2}, the tracer only counts up if a specified state transition occurs. It takes at least M cycles for the first triggering event to occur. After this transition, the shortest sequence of transition that could result in a count-up is L, where the state transition is repeated.  The number of times the triggering state transition should be visited is $C=2^w$ times. After $M+(C\times L)$ cycles, the tracer portion of the protected pattern is ready for fault generation. However, we still need another $Q$ cycles to reach a state whose encoding completes the protected pattern. If the target state could not be met in $M+(C\times L)+Q$ cycles, it might need to repeat $(C*L)$ for $N$ times to be able to reach the target state. So the number needed cycles for generating the deep fault is $M+N(C\times L)+Q$. The lower bound of required cycles (equal to the number of DIS) occurs at $N=1$, thus the minimum required BMC bound for the discovery of fault is $M+(C\times L)+Q$. $\blacksquare$

\begin{figure}
    \vspace{-10pt}
    \centering
    \includegraphics[width=0.85\columnwidth]{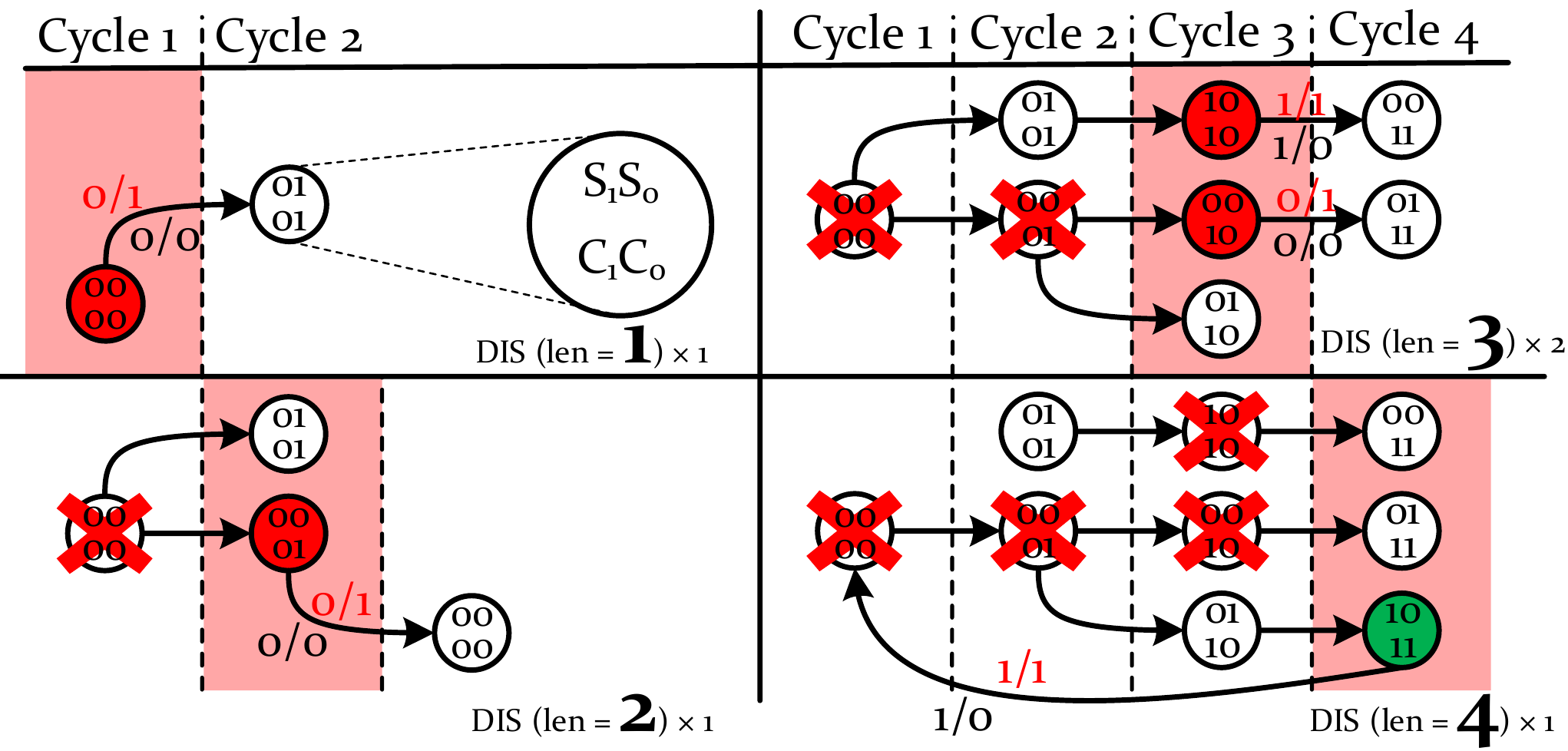}
    \caption{Cycle by cycle DIS discovery for the 011 detector of Fig. \ref{deepfault}, which is obfuscated using DF with a 2-bit counter. The deep fault is discovered when the deep state representing the protected state (shown in green) is reached.}
    \label{dis_gen}
\end{figure}

Fig.~\ref{dis_gen} shows the result of the UB-SAT attack on the 011 detector of Fig.~\ref{deepfault}. With a 2-bit counter, according to Lemma~1, the BMC min-bound of discovering faults is~$2^2=4$. As expected, in each cycle at least one fault is discovered, while the deep fault is discovered at the expected boundary of 4. 

\subsection{Preventing the Removal of the Tracer}
A simple mechanism to implement the DF-tracer is using a counter. However, counters can be easily identified by structural analysis \cite{6513710} as they have well-defined structure and are loosely connected to the rest of the circuit. Note that, for implementing the DF, the exact counter behavior is not needed; We only need a \underline {tracer circuit} for tracking cycles or events. Hence, we can use an event tracking LFSR (i.e. where LFSR state is updated based on a state transition) or a function-modified (with different encoding) counter to implement the tracer. The repetition period of LFSR (number of non-repeated LSFR state values) would serve as the depth that the fault could be delayed. In addition, to prevent the attacker from structural analysis using asynchronous signals, the enable/rest signal of the tracer circuit should not be separated from the rest of the circuit. furthermore, the tracer could be designed to exhibit a pseudo-counter behaviour such that the update of the tracer's different register state values relies on both the existing tracer register values and other registers selected from the sequential circuit or FSM.

\begin{figure}
    \centering
    \vspace{-15pt}
    \subfloat[]{{\includegraphics[width=0.27\columnwidth]{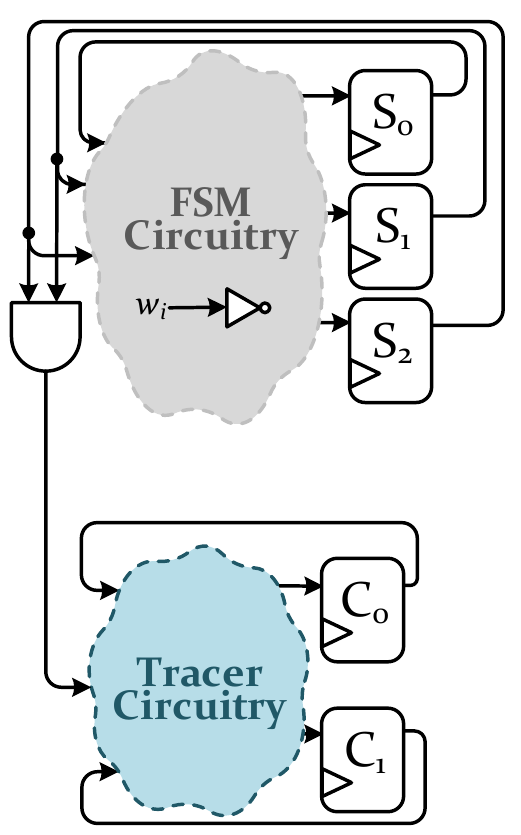}}}
    \subfloat[]{{\includegraphics[width=0.27\columnwidth]{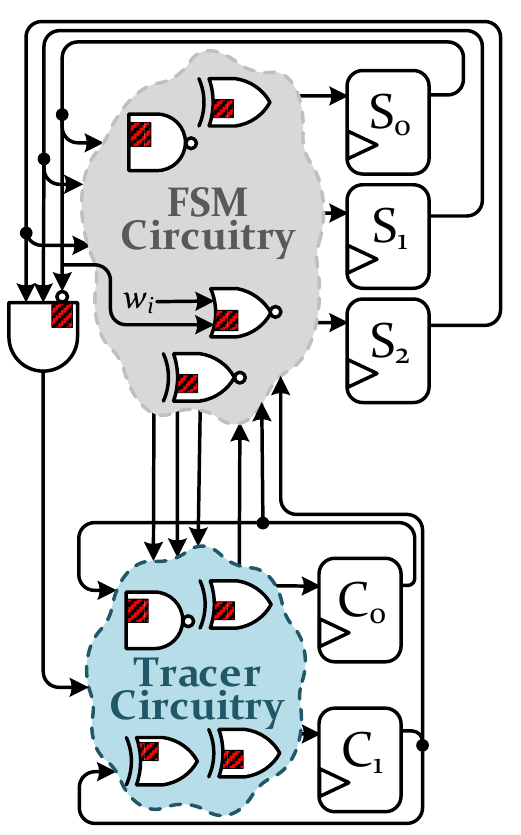}}}
    \subfloat[]{{\includegraphics[width=0.27\columnwidth]{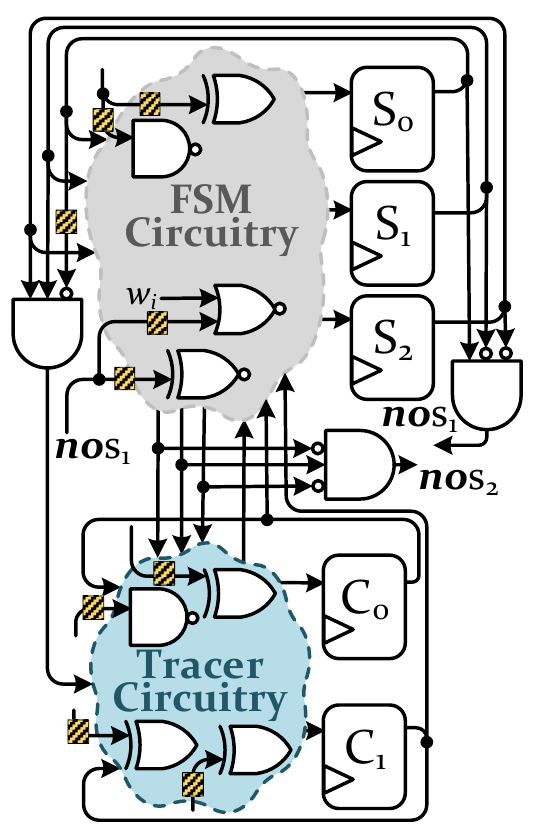}}} \\
    \vspace{-5pt}
    \subfloat[]{{\includegraphics[width=0.36\columnwidth]{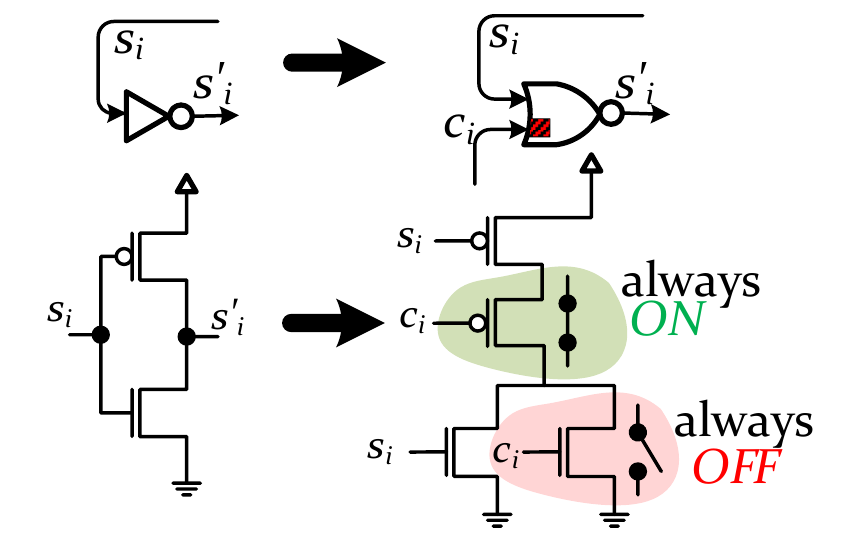}}}
    \subfloat[]{{\includegraphics[width=0.36\columnwidth]{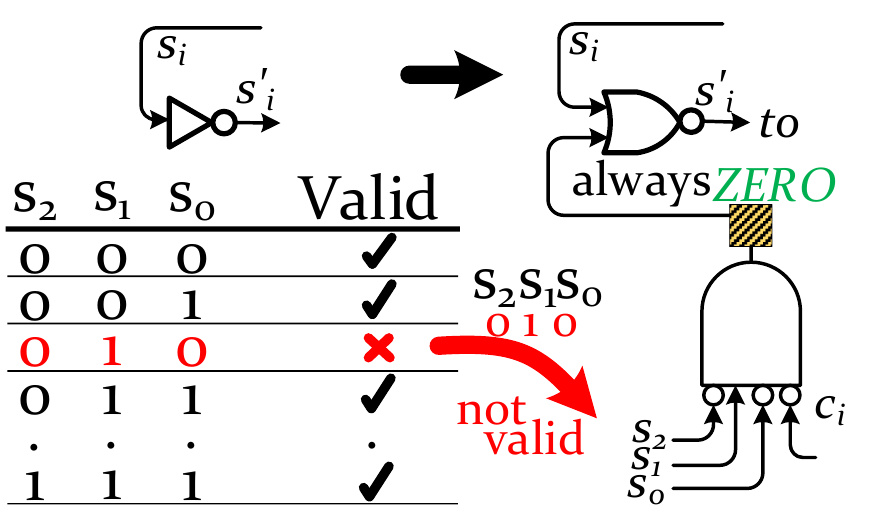}}}
    \caption{Camouflaged (Covert) gates and non-occurring states could be used for merging the two strongly-connected graphs of state and counter FFs. a)~DF circuit, b) DF circuit hidden by covert gates, c) DF circuit hidden by building dummy logic from non-occurring signal combinations, d) a covert gate implementation with dummy input, e) non-occurring signal used for implementing dummy logic.}
    \label{scc_fig}
\end{figure}

With the changes discussed earlier, the inserted tracer (modified-counter or LFSR) can not be functionally identified. However, it is still prone to detection by structural analysis of the data flow graph. As Fig. \ref{scc_fig}.(a) shows, the tracer is still loosely connected to the rest of circuit. To resolve this issue, the data flow graph of the obfuscated circuit should be modified such that tracer circuit can not be easily isolated. In other words, the tracer's registers' values should be also in the input logic cone of the other FFs. However, this should not affect the functionality of the tracer. One solution for modifying the data flow graph, as illustrated in Fig. \ref{scc_fig}.(b) is through the usage of Covert Gates \cite{Shakya_Shen_Tehranipoor_Forte_2019}. These gates have dummy inputs, connected to always on or always off transistors, that don't affect the gates' function. In practice, the gates in the logic cone of the state registers can be replaced by Covert gates, and the tracer registers' output can be connected to the dummy inputs of the covert gates. With this change, without modifying the circuit functionality, the tracer circuit will be strongly connected to the rest of the circuit when the data flow graph is extracted. The covert gates can be also used to bring additional dummy inputs from the state machine or sequential circuit to tracer without affecting its functionality. The problem with this method is that it can only protect the design against adversaries attempting to fully reverse engineer an existing ASIC, and it does not protect the IP against an untrusted manufacturing facility. In fact, the manufacturing facility has access to the layout represented via the GDSII file. Hence, the Covert gates are not hidden from the foundry.

To protect against adversarial reverse engineering at untrusted foundries, one can also utilize non-occurring signal combinations in the netlist for building dummy connections to/from the tracer circuit. As Fig. \ref{scc_fig} shows, the non-occurring signal combinations can be found using a QBF solver~\cite{shervintvlsi2020} and utilized to design an always-zero (or one) signal combined from counter FFs and signals in input cone of other state FFs. 

\begin{table*}
\scriptsize
\centering
\caption{Experimental results for shallow state duality (SSD) and deep faults (DF) methods. Iteration of last DIS (D) and its length (S) is reported in \textit{D/S} columns. Time is in seconds. Successful termination condition is reported in \textit{Term} column.}
\label{urs_result}
\setlength\tabcolsep{2pt} 
\begin{tabular}{@{} c|cccc||ccc|ccc|ccc|ccc|ccc|ccc|ccc @{}}
\toprule 
\multirow{4}{*}{Circuit} & \multicolumn{4}{c}{Circuit Info} & \multicolumn{3}{c}{Shallow State Duality} & \multicolumn{12}{c}{Deep Faults} & \multicolumn{6}{c}{DFSSD} \\
\cmidrule(lr){6-8} \cmidrule(lr){9-20} \cmidrule(lr){21-26}
 & \multicolumn{4}{c}{URS count from \cite{670866}} & \multicolumn{3}{c}{SSD} & \multicolumn{3}{c}{DF3} & \multicolumn{3}{c}{DF4} & \multicolumn{3}{c}{DF5}  & \multicolumn{3}{c}{DF7} & \multicolumn{3}{c}{SSD + DF5} & \multicolumn{3}{c}{SSD + DF7} \\
\cmidrule(lr){2-5} \cmidrule(lr){6-8} \cmidrule(lr){9-11} \cmidrule(lr){12-14} \cmidrule(lr){15-17} \cmidrule(lr){18-20} \cmidrule(lr){21-23} \cmidrule(lr){24-26}
 & FF & PI & PO & URS & D/S & Time & Term & D/S & Time & Term  & D/S & Time & Term & D/S & Time & Term & D/S & Time & Term  & D/S & Time & Term  & D/S & Time & Term \\ \cmidrule(lr){1-1} \cmidrule(lr){2-5} \cmidrule(lr){6-26}
s344   & 15   & 9  & 11  & 9,536        & 0   & 9    & UMC & 29/8  & 39   & UC  & 56/16 & 139  & UC  & 119/32 & 2849 & UC & - & TO & - & 115/32 & 5231 & UMC & - & TO  & - \\
s382   & 21   & 3  & 6   & 2,073,412    & 0   & 5    & UMC & 17/48     & 9393  & UMC   & -     & TO   & -   & -      & TO   & -   & - & TO & - & -      & TO   & -   & - & TO  & - \\
s386   & 6    & 7  & 7   & 51           & 0   & 6    & UMC & 8/8   & 9    & UC  & 29/16 & 45   & UC  & 70/32  & 482  & UC & - & TO & - & 65/32  & 1232 & UMC & - & TO  & - \\
s526   & 21   & 3  & 6   & 1,695,692    & 0   & 5    & UMC & 27/16 & 463  & UMC & 49/32 & 1580 & UMC & 92/64  & 4684 & UMC & - & TO & - & 90/64  & 4676 & UMC & - & TO  & - \\
s713   & 19   & 35 & 23  & 517,625      & 0   & 223  & UMC & 15/8  & 28   & UC  & 53/16 & 317  & UC  & -      & TO   & -   & - & TO & - & -      & TO   & -   & - & TO  & - \\
s832   & 5    & 18 & 19  & 7            & 0   & 12   & UMC & 32/16 & 86   & UC  & 44/16 & 148  & UC  & 99/32  & 2549 & UC & - & TO & - & -      & TO  & -   & - & TO & - \\
s838   & 32   & 34 & 1   & $>2^{30}$    & -   & TO   & -   & 20/72 & 1096 & UC & 43/80 & 2194 & UC  & -      & TO   & -   & - & TO & - & -      & TO   & -   & - & TO  & - \\
s1196  & 18   & 14 & 14  & 259,492      & 0   & 12   & UMC & 27/8  & 77   & UC  & 64/16 & 377  & UC  & 135/32 & 3146 & UC & - & TO & - & -      & TO   & -   & - & TO  & - \\
s1423  & 74   & 17 & 5   & $>2^{72}$    & -   & TO   & -   & 8/8   & 26   & UC  & 49/16 & 2924 & UC  & -      & TO   & -   & - & TO & - & -      & TO   & -   & - & TO  & - \\
s1494  & 6    & 8  & 19  & 16           & 0   & 15   & UMC & 21/8  & 46   & UC  & 78/16 & 369  & UC  & 187/32 & 4333 & UC  & - & TO & - & -      & TO   & -   & - & TO  & - \\
s5378  & 179  & 35 & 49  & $>2^{176}$   & 0   & 65   & UMC & 42/8  & 1653 & UC  & -     & TO   & -   & -      & TO   & -   & - & TO & - & -      & TO   & -   & - & TO  & - \\
s9234  & 211  & 36 & 39  & $>2^{227}$   & 0   & 3065 & UMC & -     & TO   & -   & -     & TO   & -   & -      & TO   & -   & - & TO & - & -      & TO   & -   & - & TO  & - \\
s38584 & 1426 & 38 & 304 & $>2^{1449}$  & -   & TO  & -   & 8/8   & 1293 & UC  & -     & TO   & -   & -      & TO   & -   & - & TO & - & -      & TO  & -   & - & TO & - \\
\bottomrule
\end{tabular}
\end{table*}

\section{Experimental Results} \label{results}
We have implemented the UB-SAT attack using Yices SMT solver by creating the combinational equivalent circuit and unrolling it for finding DISes and checking UC and CE terminations. For UMC termination, we have used SuperProve from Berkeley ABC package \cite{sterin2011benefit}. The experiments were performed on an Intel~Core~i5 with 64GB RAM.

Table \ref{urs_result} captures the results of attacking the ISCAS'89 benchmarks when encoded using duplicated states (SSD), obfuscated using deep faults (DF), protected using a combination of both techniques (DFSSD). The first few columns of the table describe the characteristics of these benchmarks in terms of number of flip-flops (FF), number of primary I/O (PI and PO), and number of unreachable states (URS) according to \cite{670866}. In this table DF$w$ represent a DF obfuscation, constructed using a counter of width~$w$. For each obfuscated circuit, number of discovered DISes, and the number of inputs in the last DIS (its length) is reported as D/S. The maximum attack time is set to eight hours. Attacks that take longer are reported as \textit{TO}.

The SSD column of Table \ref{urs_result} captures the result of UB-SAT attack against circuits protected only by Shallow State Duality. As expected, when UB-SAT is deployed against a SSD encoded circuit, the UC And CE termination strategies become useless. As reported, for all SSD-encoded benchmarks, either the attack is terminated by UMC or prematurely terminated for lack of memory resources. 

Table \ref{urs_result} also captures the results of attacking circuits, which are obfuscated using deep faults with varying counter widths (3, 4, 5, and 7 bits). From this table, following observations are made: 1) the number of discovered DISes grow exponentially with respect to the size of the counter. This is consistent with the Lemma 1: at each cycle we can produce at least 1 DIS until the protected pattern (which in this case is encoded using the highest value of the counter) is reached. Hence, we should at least have $2^w$ DISes. 2) The size of the largest input sequence (S) in which the deep fault is discovered is $N\times 2^w$ (N being an integer). This is consistent with  Lemma 1, where the protected pattern could be potentially (but not necessarily) observed at every N$\times 2^w$ cycles; 3) the runtime of the attack increases exponentially as the depth of DF tracer circuit (counter) increases; and 4) when the circuit is solely protected by DF, the UC termination is the most reoccurring termination strategy.

The last two columns of Table \ref{urs_result}, capture the impact of combining the DF and SSD (DFSSD) which is the main solution proposed in this paper. The DFSSD combines the best feature of the two solutions. The SSD prevents early UC and CE termination, while the DF pushes the faults down into deep states, resulting in an exponential increase in the number of required DISes and the attack time with respect to the counter size. Note that by preventing the UC and CE terminations, and by forcing the attack to UMC termination check in every iteration, the SSD+DF5 has considerably larger runtime compared to DF5.

\section{Conclusion}
In this paper, we proposed DFSSD, an obfuscation solution for FSM and sequential circuits with restricted (locked) access to the scan chain. The DFSSD deployed two mechanisms, specifically designed to resist against BMC-based attacks such as UB-SAT: 1) it uses shallow state duality to prevent early termination of such attacks by invalidating the unique completion and combinational equivalence checks, forcing the attack to rely on exhaustive and time-consuming UMC for assessing the attack's termination condition; 2) it injects fault into deep and hard to reach (by a BMC) states. The DFSSD allows the designer to precisely control the depth of the fault at design time using a low overhead circuit solution and make the attack time unreasonably long. 

\section{Acknowledgement}
This research is funded by the Defense Advanced Research Projects Agency (DARPA \#FA8650-18-1-7819) of the USA, and partly by Silicon Research Co. (SRC TaskID 2772.001) and National Science Foundation (NSF Award\# 1718434).

\renewcommand{\IEEEbibitemsep}{0pt plus 0.5pt}
\makeatletter
\IEEEtriggercmd{\reset@font\normalfont\fontsize{6.0pt}{6.8pt}\selectfont}
\makeatother
\IEEEtriggeratref{1}

\bibliographystyle{IEEEtran}
\bibliography{IEEEabrv,refs}

\end{document}